\documentclass[aps,pra,twocolumn,groupedaddress,showpacs]{revtex4}
\usepackage{amsmath}
\usepackage{graphicx}
\usepackage{amsfonts}
\usepackage{color}
\makeatletter
\def\captionof#1#2{{\def\@captype{#1}#2}}
\makeatother
\begin{document}

\title{Out of equilibrium open quantum systems: a comparison of approximate Quantum Master Equation approaches with  exact results} %

\author{Archak Purkayastha}
\affiliation{International centre for theoretical sciences, Tata Institute of Fundamental Research, Bangalore - 560012, India}
\author{Manas Kulkarni}
\affiliation{Department of Physics, New York City College of
Technology, The City University of New York, Brooklyn, NY 11201, USA}
\author{Abhishek Dhar}
\affiliation{International centre for theoretical sciences, Tata Institute of Fundamental Research, Bangalore - 560012, India}

\begin{abstract}
We present the Born-Markov approximated Redfield quantum master equation (RQME) description for an open system of non-interacting particles (bosons or fermions) on an arbitrary lattice of $N$ sites in any dimension and weakly connected to multiple reservoirs at different temperatures and chemical potentials. The RQME can be reduced to the Lindblad equation, of various forms, by making further approximations. By studying the $N=2$ case, we show that RQME gives results which agree with exact analytical results for steady state properties and with exact numerics for time-dependent properties, over a wide range of parameters. In comparison, the Lindblad equations have a limited domain of validity in non-equilibrium. We conclude that it is indeed justified to use microscopically derived full RQME to go beyond the limitations of Lindblad equations in out-of-equilibrium systems. We also derive closed form analytical results for out-of-equilibrium time dynamics of two-point correlation functions. These results explicitly show the approach to steady state and thermalization. These results are experimentally relevant for  cold atoms, cavity QED and far-from-equilibrium quantum dot experiments.  
\end{abstract}
\maketitle
\section{Introduction}
\begin{figure}[!t]
\includegraphics[scale=.34]{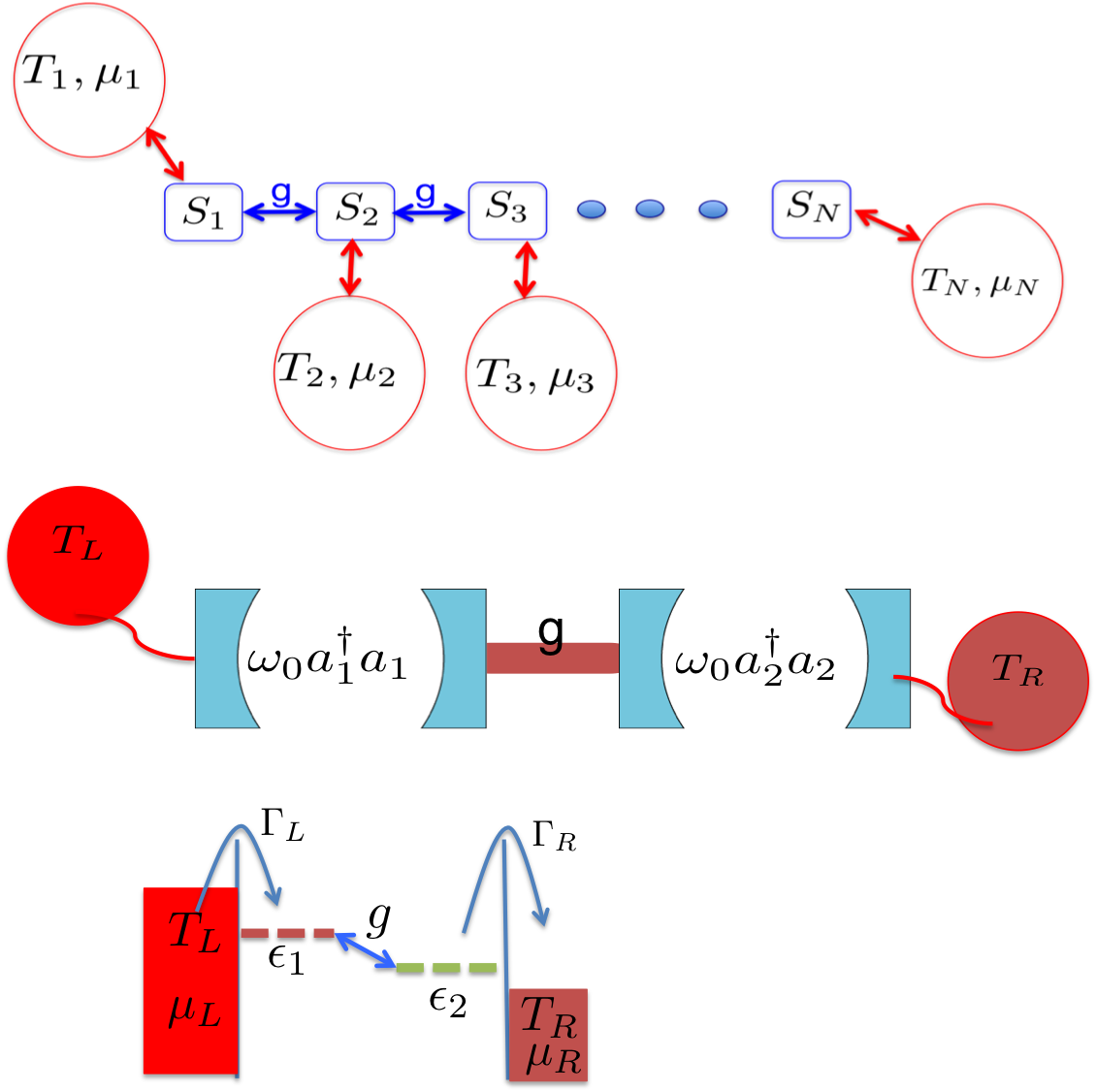} 
\caption{\footnotesize (color online) (Top) A schematic of the setup we consider. We have a system of ``$N$'' lattice sites with non-interacting particles on them. Each site of the system is coupled to its respective bath of a chosen temperature and chemical potential. Baths out of equilibrium can facilitate transport. The above diagram is general for bosonic or fermionic case and such situations have been realized experimentally in setups such as given in middle and bottom figures.} \label{fig:zero}
\end{figure}

Understanding out-of-equilibrium quantum systems, both bosonic and fermionic has  been of great experimental and theoretical interest recently. Such open quantum systems have been realized both in the bosonic (experiments on cavity-QED arrays \cite{KLH2015, nat_phys_rev,underwood_PhysRevA.86.023837,sskj}, cold atoms \cite{esslinger10, kulkarniprl}, cavity optomechanics \cite{cavop,cavop1}) and fermionic case (non-equilibrium transport in coupled quantum dots \cite{KLH2015,Kulkarni2014,Liu2014,Agarwalla2016,Hartle2015}, cold atoms \cite{Krinner2015}). A typical setup of interest (see Fig.~\ref{fig:zero}) is a system of small number of bosonic (fermionic) degrees of freedom coupled to a bosonic (fermionic) bath/environment composed of large number of degrees of freedom. The general case can be defined as a scenario in which each degree of freedom of the sub-system is coupled to a bath characterized by a temperature ($T={1}/{\beta}$) and a chemical potential ($\mu$). New cutting edge technologies available recently for measuring physical quantities such as occupation number, currents and correlations in such  systems makes it of paramount importance to develop an approach that produces accurate results.  Additionally, these open quantum systems are widely tunable thereby providing a large window of parameters to test validity of different approaches.

One of the most commonly-used frameworks in the study of open quantum systems in the limit of weak system-bath coupling is the quantum master equation (QME) method, where one writes a time-evolution equation for the reduced density matrix $\rho$ of the system. Two  QMEs' that are popular in the literature  are the Redfield \cite{Redfield1965} and the Lindblad equations \cite{Lindblad1976,Gorini1976}. The Redfield equations (RQME) are obtained under the so-called Born-Markov approximation (see later).  After further approximations, and depending on the nature of these, we get either the local coordiante basis (site-basis) Lindblad QME (LLQME) or  the eigenfunction basis Lindblad QME (ELQME).  An advantage of the Lindblad equations is that they have been proven to preserve positivity of $\rho$. Hence they are widely used in the literature \cite{Torres2014,Esposito2007,Ajisaka2012,Prosen2008,Prosen2010,Znidaricjpa,Ajisaka2015,
GSA2013,Asadian2013, BauerNotes, Rivas2011, Levy2014, Wichterich2007}. However, both Lindblad equations are known to have limitations. The local basis Lindblad does not give the correct thermalization \cite{BauerNotes,Levy2014} for non-interacting bosons/fermions, while the eigenbasis Lindblad gives zero particle current inside the system even in the non-equilibrium steady state (NESS) \cite{BauerNotes, Wichterich2007}. 

The RQME provides an alternative. It has been previously used to find NESS properties in open quantum harmonic chains, spin chains and fermionic lattices \cite{rep1, Wu2010,Pepino2010,Saito2000,Saito2003,Wu2011,Harbola2006,Kondov2001,Kondov2003}. RQME is known to give thermalization, but is more difficult to solve, both numerically and analytically. Also, it does not guarantee the complete positivity of the system density matrix at all times. This may lead to various subtle pathologies depending on initial conditions, as shown in some recent works \cite{Anderloni2007,Argentieri2014}. The Lindblad equations, on the other hand, are free from these pathologies. 

Because of such inherent pathologies and limitations stemming from the approximations,  it is important to check performance of the approximate QMEs against exact results. Results from RQME, Lindblad QMEs and exact calculations have been recently compared for a single oscillator in NESS \cite{Thingna2013} and for a two site bosonic problem in equilibrium \cite{Keeling2015}. To our knowledge, rigorous checks have not been performed for multiple sites in non-equilibrium. 

In this manuscript, we ask the question whether it is justified to adopt RQME to go beyond the limitations of the Lindblad methods in non-equilibrium systems. To this end, first, for an out-of-equilibrium system of non-interacting bosons/fermions in an arbitrary lattice of $N$ sites in any dimension, we derive a closed set of linear differential equations for two-point correlation functions from RQME. Such closed set of linear differential equations can be easily solved numerically. This, therefore, gives a numerical way to compute physical results for such out-of-equilibrium systems from RQME. Next, to check validity of such solutions, we compute physical quantities of interest from RQME for a system of two bosonic/fermionic modes connected to two baths at different temperatures and chemical potentials (i.e, out-of-equilibrium) and compare them against exact results obtained from other open system approaches, such as the quantum Langevin method (QLE), where exact equations for system degrees of freedom admit the form of an effective generalized `Langevin equation', and equation of motion (EOM) method \cite{fordetal1965,dharroy2006,dharsen2006,dharetal2012}. We find that RQME indeed reproduces the exact results quite accurately. We also provide closed form analytical results for out-of-equilibrium time dynamics of two-point correlation functions for the two site problem. Throughout the paper, $\hbar$ is taken to be unity.

The plan of the paper is as follows. In Sec.~(\ref{model}) we define the general model and give a derivation of the Redfield QME. In Sec.~(\ref{comparison}) we consider a two site example and compare results obtained from  different QMEs' with  exact results. Finally we summarize our results in Sec.~(\ref{conclusions}).

\section{The model and the Redfield QME (RQME)}
\label{model}
{\bf Definition of model}: We consider non-interacting bosonic (fermionic) particles on a lattice of $N$ sites in arbitrary dimension and of arbitrary geometry where each site  is coupled to  bosonic (fermionic) baths. The bath Hamiltonian, $\hat{\mathcal{H}}_B$, as well as the coupling between system and baths, $\hat{\mathcal{H}}_{SB}$, are taken to be bilinear. The full Hamiltonian thus takes the form
\begin{align}
\label{model_H}
\hat{\mathcal{H}} &= \hat{\mathcal{H}}_S + \hat{\mathcal{H}}_B + \hat{\mathcal{H}}_{SB}~,~~{\rm where}  \nonumber\\
~\hat{\mathcal{H}}_S &= \sum_{\ell=1}^N H^{(S)}_{\ell m} \hat{a}_\ell^{\dagger} \hat{a}_m~,~~~
\hat{\mathcal{H}}_B = \sum_{\ell=1}^N\sum_{r=1}^\infty \Omega_r^\ell \hat{B}_r^{\ell \dagger} \hat{B}_r^\ell~, \nonumber\\
\hat{\mathcal{H}}_{SB} &= \varepsilon\sum_{\ell=1}^N \sum_{r} (\kappa_{\ell r} \hat{B}_r^{\ell \dagger} \hat{a}_\ell + \kappa_{\ell r}^* \hat{a}_\ell^{\dagger} \hat{B}_r^\ell)~,   
\end{align} 
where $H^{(S)}$ is a Hermitian matrix and $\hat{a}_\ell$ correspond to bosonic (fermionic) annihilation operators defined respectively on $\ell$th lattice point of system and  $\hat{B}_{r}^{\ell}$ to those of baths attached to the $\ell$th point.  
The baths  have infinite degrees of freedom. $\varepsilon$ is a parameter that controls system bath coupling and has dimensions of energy, so that $\{\kappa_{\ell r}\}$ are dimensionless and $O(1)$. We assume that, initially, each of the $\ell$ baths is at thermal equilibrium at its own inverse temperature $\beta_\ell$ and chemical potential $\mu_\ell$, and there is no coupling between system and baths. Thus, the  initial bath correlation functions satisfy the thermal properties :
\begin{equation}
\label{initial_bath_corr}
\begin{split}
\langle\hat{B}_r^\ell\rangle = 0 , \hspace{2pt} <\hat{B}_r^{\ell\dagger} \hat{B}_s^\ell >_B = n_{\ell}(\Omega_r^\ell)\delta_{r s }~,  
\end{split}
\end{equation}
where $n_\ell(\omega)= [{e^{\beta_{\ell}( \omega-\mu_{\ell})}\pm 1}]^{-1}$ is the fermionic or bosonic distribution function. We also introduce the bath spectral functions:
\begin{equation}
\label{J}
J_{\ell}(\omega)=2\pi \sum_r  \mid \kappa_{\ell r} \mid^2 \delta(\omega - \Omega_r^\ell) ~.
\end{equation}  

{\bf Redfield QME}: 
In the Redfield approach, one assumes  weak system-bath coupling limit. Performing the Born-Markov approximation leads  to the standard Redfield equation \cite{CarmichaelBook,GSAbook}. 
For bilinear system-bath coupling $\hat{\mathcal{H}}_{SB} = \sum_{\ell} \hat{S}_\ell \hat{B}_\ell$
where $\hat{S}_\ell$ operates on system and $\hat{B}_\ell$ operates on bath, system-bath coupling being turned on at time $t=0$, we have the master equation 
\begin{align}
\label{gen_master}
\frac{\partial\rho^I}{\partial t} &= -\sum_{\ell,m} \int_0^t dt^{\prime} \Big\{ [\hat{S}^I_\ell(t),\hat{S}^I_m(t^{\prime})\rho^I(t)]\langle \hat{B}^I_\ell(t)\hat{B}^I_m(t^\prime)\rangle_B  \nonumber \\ & +[\rho^I(t)\hat{S}^I_m(t^{\prime}),\hat{S}^I_\ell(t)]\langle \hat{B}^I_m(t^\prime)\hat{B}^I_\ell(t)\rangle_B \Big\}~, 
\end{align}
where we use the interaction representation $\hat{O}^I(t)=e^{i(\hat{\mathcal{H}}_S+\hat{\mathcal{H}}_B)t} \hat{O} e^{-i(\hat{\mathcal{H}}_S+\hat{\mathcal{H}}_B)t}$ and  $\langle...\rangle_B$ refers to the average taken only with respect to bath. In order to obtain the explicit QME for our model, it is convenient to go to the eigenbasis  of the system Hamiltonian. Let $c$ be the unitary matrix which diagonalizes $H^{(S)}$, i.e., 
\begin{equation}
c^\dagger H^{(S)}  c= \omega^{(D)}~,
\end{equation}
where  $c^\dagger c=I$ and $\omega^{(D)}$ is diagonal matrix with elements $\omega_\nu$.  Then we also define new operators $\{A_\alpha\}$ through the transformation \begin{equation}
\hat{a}_{\ell} = \sum_{\alpha=1}^N c_{\ell \alpha} \hat{A}_\alpha~.
\label{Aop}
\end{equation}
 Thus $\hat{A}_\alpha$ is the annihilation operator for the $\alpha$th eigen-mode with energy $\omega_\alpha$. For RQME to be valid, we choose $\varepsilon \ll \{\omega_{\alpha} \}$. After some tedious algebra (see Appendix), we obtain our QME for the reduced density matrix $\rho$ of the system:
\begin{align}
&\frac{\partial\rho}{\partial t} = i[\rho, H_S] \nonumber\\
&-\varepsilon^2 \sum_{\alpha,\nu=1}^N \int \frac{d\omega}{\pi}\Big[\int_0^\infty d\tau e^{i(\omega - \omega_\nu)\tau}\mathcal{L}(\hat{A}_\alpha^{\dagger},\hat{A}_\nu;\omega)\rho + \hspace{2pt} h.c.\Big]~,\nonumber \\
&{\rm where}
\label{qme} \\
&\mathcal{L}(\hat{A}_\alpha^{\dagger},\hat{A}_\nu;\omega)\rho =(f_{\alpha\nu}(\omega) \mp F_{\alpha\nu}(\omega))[\hat{A}_\alpha^{\dagger}, \hat{A}_\nu \rho] \nonumber \\   
&+ F_{\alpha\nu}(\omega)[\rho \hat{A}_\nu, \hat{A}_\alpha^{\dagger}], \nonumber \\
&f_{\alpha\nu}(\omega) = \sum_{{\ell}=1}^Nc_{\ell \alpha}^*c_{\ell \nu} \frac{J_{\ell}(\omega)}{2}, \hspace{5pt}
F_{\alpha\nu}(\omega) = \sum_{{\ell}=1}^N c_{\ell \alpha}^*c_{\ell \nu} \frac{J_{\ell}(\omega)n_{\ell}(\omega)}{2}~. \nonumber
\end{align}
Here the integration over $\omega$ is over all bath energy levels. Also we have taken observation times $t \gg \tau_B$, where $\tau_B$ is the characteristic relaxation time scale of the bath. For baths with wide bandwidth and at low temperature, $\tau_B \sim \beta$(see Appendix). Similar RQME was dervied in Ref.~\cite{Wu2010,Pepino2010} for slightly different systems.   

{\bf Equations for correlation functions}:~We note that it is possible to obtain closed time-depedent equations for the full set of two point correlations $C_{\alpha\nu}(t) = Tr(\rho(t) \hat{A}_{\alpha}^{\dagger}\hat{A}_{\nu})$ .
The evolution equation for $C_{\alpha\nu}(t)$ as obtained from the QME is :
\begin{align}
\label{C_differential}
&\frac{dC_{\alpha\nu}}{dt} = i\omega_\alpha C_{\alpha\nu}(t) + \varepsilon^2 \Big [ F_{\nu\alpha}(\omega_\alpha)-i F_{\nu\alpha}^\Delta(\omega_\alpha) \nonumber \\
&- \sum_{\sigma=1}^N C_{\alpha\sigma}(t) v_{\nu \sigma }\Big ] 
+(\alpha \leftrightarrow \nu )^\dagger ~,~~{\rm where} \nonumber \\ 
& v_{\alpha \nu}=f_{\alpha\nu}(\omega_{\nu})-if_{\alpha\nu}^{\Delta}(\omega_{\nu})~,
\end{align}
with $f_{\alpha\nu}^{\Delta}(\omega) = \mathcal{P} \int \frac{}{}\frac{d\omega^{\prime} f_{\alpha\nu}(\omega^{\prime})}{\pi(\omega^{\prime}-\omega)}, \hspace{2pt}
F_{\alpha\nu}^{\Delta}(\omega) = \mathcal{P} \int \frac{}{}\frac{d\omega^{\prime} F_{\alpha\nu}(\omega^{\prime})}{\pi(\omega^{\prime}-\omega)}$, where $\mathcal{P}$ denotes principal value. 

The Redfield equation is, in general, quite difficult to solve. That is one of the main reasons for reduction of Redfield equation to Lindblad forms, for which various numerical techniques are available. However, note that Eq.~\ref{C_differential} forms a closed set of linear differential equations which can be easily solved numerically. Various physical quantities of interest such as occupation density and particle current can be obtained from these correlation functions. The main pre-requisite for solving Eq.~\ref{C_differential} is diagonalization of the system Hamiltonian to go to the eigenbasis. For a non-interacting bosonic/fermionic system of $N$ sites in arbitrary lattices, this only requires diagonalization of the $N \times N$ matrix $H^{(S)}$. If there are no further symmetries in the system, one needs all the $N^2$ two-point correlation functions to close the set of linear differential equations, and hence needs to deal with $N^2 \times N^2$ matrices. Thus, Eq.~\ref{C_differential} gives a numerical way to directly calculate time-dynamics of out-of-equilibrium two-point correlation functions in an arbitrary system of non-interacting particles in a lattice. In the rest of the manuscript we will be testing the validity of solutions of Eq.~\ref{C_differential}.

\section{$N=2$ case: Comparison between QME and exact results}
\label{comparison}

The solution of Eq.~\ref{C_differential} gives two point correlation functions for the $N$ site non-interacting bosons/fermions in non-equlibrium in any dimension. We propose to check the validity of such solutions. To do so,  we now go to a specific simple problem with $N=2$ sites and check how good Eq.~\ref{C_differential} does when compared to exact results.

We consider the following specific two-site system  coupled to baths which are one-dimensional chains:
\begin{align}
\hat{\mathcal{H}}_S &= \omega_{0} (\hat{a}_1^{\dagger} \hat{a}_1+\hat{a}_2^{\dagger} \hat{a}_2) + g(\hat{a}_1^{\dagger} \hat{a}_{2} + \hat{a}_{2}^{\dagger} \hat{a}_1)~, \nonumber \\
\hat{\mathcal{H}}_B^{(\ell)}&= t_B (\sum_{s=1}^\infty \hat{b}_s^{\ell\dagger}\hat{b}_{s+1}^\ell+h.c.), \hspace{5pt} \hat{\mathcal{H}}_B = \hat{\mathcal{H}}_B^{(1)}+\hat{\mathcal{H}}_B^{(2)}~,\nonumber \\
\hat{\mathcal{H}}_{SB} &= \varepsilon\gamma_1 (\hat{a}_1^{\dagger}\hat{b}_{1}^1+ h.c.) + \varepsilon\gamma_2 (\hat{a}_2^{\dagger}\hat{b}_{1}^2+ h.c.)~,  \label{ham2S} 
\end{align}
where the operators are either all bosonic or all fermionic and $\hat{b}_{s}^\ell$ is the annihilation operator of the $s$th bath site of the $\ell$th bath. The eigenmodes of the system are given by $\hat{A}_1 = {(\hat{a}_1-\hat{a}_2)}/{\sqrt{2}}$, $\hat{A}_2 = {(\hat{a}_1+\hat{a}_2)}/{\sqrt{2}}$ with eigenvalues $\omega_1 = \omega_0 - g$, $\omega_2 = \omega_0 + g$. We assume $\omega_0 \gg \varepsilon$, so that QME can be applied, while the parameter $g$ can be varied freely. The bath spectral functions, defined in Eq.~\ref{J}, can be obtained explicitly by going to eigenmodes of the baths and are given by  \cite{dharetal2012} (see Appendix). 
\begin{equation}
J_\ell(\omega) = \frac{2\gamma_\ell^2}{t_B}\sqrt{1-\left(\frac{\omega}{2t_B}\right)^2}~.
\label{Jform}
\end{equation}

In the following sections we first give some details of the exact approaches 
to obtain steady state [Sec.~(\ref{QLE})] and time-dynamics [Sec.~(\ref{dynamics}), then present some analytic results from RQME [Sec.~(\ref{RQME})], discuss the reduction to the Lindblad form [Sec.~(\ref{lindblad})]  and finally present the comparisions between the various methods [Sec.(\ref{comp})].

\subsection{Exact results : steady state via Quantum Langevin Equation (QLE)} 
\label{QLE}
Exact steady state properties of the system can be found using QLE, as done in Ref \cite{dharsen2006, dharroy2006, dharetal2012}. We briefly outline the procedure for computing physical quantities from the QLE method for the 2-site case with Hamiltonian given by Eq.~\ref{ham2S}. We first go to the eigenmodes of the bath by doing a unitary transformation. So, $\hat{\mathcal{H}}_B^{(\ell)}= t_B (\sum_{s=1}^\infty \hat{b}_s^{\ell\dagger}\hat{b}_{s+1}^\ell+h.c.) = \sum_{r=1}^\infty \Omega_r^\ell \hat{B}_r^{\ell \dagger} \hat{B}_r^\ell$ where $\hat{b}_s^{\ell} = \sum_r U_{rs}^{\ell *} \hat{B}^\ell_r$ and $U$ is a unitary matrix that diagonalizes $\hat{\mathcal{H}}_B^{(\ell)}$. $\hat{B}^\ell_r$ is annihilation operator of the eigenmode with eigenvalue $\Omega_r^\ell$. The bath eigenmodes satisfy the initial bath correlation functions: $\langle\hat{B}_r^\ell\rangle = 0$ and $\langle\hat{B}_r^{\ell\dagger} \hat{B}_s^\ell \rangle = n_{\ell}(\Omega_r^\ell)\delta_{r s }$. We then have $\kappa_{\ell r} = \gamma_\ell U_{r1}^{\ell *}$. The equation of motion (EOM) for the system and the bath are 
\begin{align}
&\frac{d{\hat{B}}^\ell_r}{dt} = -i[\hat{B}^\ell_r,\hat{\mathcal{H}}] = -i(\Omega^\ell_r \hat{B}^\ell_r + \varepsilon\kappa_{\ell r}^* \hat{a}_\ell)~, \label{bath_eom}\\
&\frac{d\hat{a}_1}{dt} = -i(\omega_0 \hat{a}_1 + g \hat{a}_2 + \sum_{r} \varepsilon\kappa_{1 r}\hat{B}^{(1)}_r )\nonumber \\& \hspace{5pt} (1\leftrightarrow 2)~. \label{sys_eom}
\end{align}
The equation for the bath (Eq.~\ref{bath_eom}) can be formally solved to obtain :
\begin{equation}
\hat{B}^\ell_r(t) = \hat{B}^\ell_r(0)e^{-i\Omega^\ell_r t} -i \varepsilon \kappa_{\ell r}^* \int_0^t dt^\prime \hat{a}_\ell(t^{\prime})e^{-i\Omega^\ell_r (t-t^\prime)}~. 
\end{equation}
Using this in Eq.~\ref{sys_eom}, we obtain the quantum Langevin equation (QLE)
\begin{align}
&\frac{d\hat{a}_1}{dt} = -i(\omega_0 \hat{a}_1 + g \hat{a}_2) -i \varepsilon \hat{\xi}_1 - \varepsilon^2 \int^t_0 dt^{\prime} \hat{a}_1(t^{\prime}) \alpha_1(t-t^{\prime}), \nonumber \\& \hspace{5pt} (1\leftrightarrow 2)~, \label{qleEOM}
\end{align}
where
\begin{align}
\hat{\xi}_\ell(t) &= \sum_r \kappa_{\ell r} \hat{B}^\ell_r(0) e^{-i\Omega^\ell_r t}~, \\
\alpha_\ell(t-t^{\prime}) &= \int \frac{d\omega}{2\pi}  J_\ell(\omega) e^{-i\omega(t-t^{\prime})}~,
\end{align} 
represent `noise' and `dissipation' respectively and $J_\ell(\omega)$ are the bath spectral functions as defined in Eq.~\ref{J}. Using the bath correlations, we obtain 
\begin{equation}
\langle\hat{\xi}^{\dagger}_{\ell}(t^{\prime}) \hat{\xi}_{\ell}(t)\rangle = \int \frac{d\omega}{2 \pi} J_\ell(\omega)n_\ell(\omega) e^{-i\omega (t-t^{\prime})}~.
\end{equation}
Note that even though Eq.~\ref{qleEOM} has the form of a generalized Langevin equation, it is completely exact. The `noise' and `dissipation' that appear in the QLE arise from exact treatment of the bath. The steady state results can be easily found from Eq.~\ref{qleEOM} by taking $t\rightarrow \infty$ and doing Fourier transforms.
The NESS current and occupation obtained are given by 
\begin{align}
I =& g^2 \varepsilon^2 \int \frac{d\omega}{2\pi} \frac{J_1(\omega)J_2(\omega)\big[n_1(\omega) - n_2(\omega)\big]}{\mid \mathcal{M}(\omega) \mid ^2}~, \label{Iqle}\\
\langle\hat{a}_1^{\dagger} \hat{a}_1\rangle=& \varepsilon \int \frac{d\omega}{2\pi} \Big[ \frac{\mathcal{K}(\omega) J_1(\omega)n_1(\omega)}{\mid \mathcal{M}(\omega) \mid ^2} 
+ \frac{g^2 J_2(\omega)n_2(\omega)}{\mid \mathcal{M}(\omega) \mid ^2} \Big]~,
\label{occqle}
\end{align}
where
\begin{align} 
&\Delta_\ell(\omega) = \mathcal{P} \int \frac{d\omega^{\prime}}{2\pi} \frac{J_\ell(\omega^{\prime})}{\omega - \omega^{\prime}}~,  \nonumber\\
&\mathcal{M}(\omega)= \Big[\big(\omega_0 - \omega - i\varepsilon^2\frac{J_1(\omega)}{2}+\varepsilon^2\Delta_1(\omega)\big) \nonumber \\
&\big(\omega_0 - \omega - i\varepsilon^2\frac{J_2(\omega)}{2}+\varepsilon^2\Delta_2(\omega)\big) - g^2 \Big]~, \nonumber \\
&\mathcal{K}(\omega)=\left| \omega_0 - \omega - i\varepsilon^2\frac{J_2(\omega)}{2}+\varepsilon^2\Delta_2(\omega) \right| ^2~. 
\end{align}
The occupation of the second site is just $<a_2^{\dagger} a_2> = 1 \leftrightarrow 2$ in Eq.~(\ref{occqle}). All integrals over $\omega$ go over all possible values of $\omega$. Note that Eq.~(\ref{Iqle}),(\ref{occqle}) are exact results without any approximation. However, they are not closed form results and involve some complicated integrals. Also, obtaining exact transient behaviour by this method is difficult. But transient behaviour can be easily obtained by exact numerics.

\subsection{Exact results : time dynamics via system+bath numerics}
\label{dynamics}
To check the time dynamics we do numerical simulations. For this purpose, we choose a bath of finite size and evolve the full system+bath Hamiltonian $\hat{\mathcal{H}}$ using unitary Hamiltonian dynamics. Let us collectively denote by ``$d$'' a column vector with all annihilation operators of both system and baths. The full Hamiltonian can be written as $\hat{\hat{\mathcal{H}}}=\sum_{i,j}H_{i j} d^\dagger_i d_j$ where $i$ now refers to either system or bath sites. If  
$ D= \langle d d^{\dagger} \rangle$ denotes the full correlation matrix of system and baths, its time evolution is given by $D(t)=e^{i H t} D e^{-i H t}$. 
In our simulations we considered the system described by Eq.~(\ref{ham2S}) of two sites connected to baths each with $511$ sites which are large enough to show negligible finite size effects.

\subsection{Correlation functions from RQME}
\label{RQME}
\subsubsection{Sysmmetric coupling to baths : $\gamma_1 = \gamma_2$: Full solution :} 

In Eq.~\ref{C_differential}, this corresponds to the special case of all system bath couplings being equal, i.e, when, $J_{\ell}(\omega)=J(\omega)$. Under this condition Eq.~\ref{C_differential} can be solved exactly using the fact that
$f_{\nu \sigma}(\omega) = [{J(\omega)}/{2}]\sum_{r=1}^Nc_{\ell \nu}^*c_{\ell \sigma} = 
[{J(\omega)}/{2}] \delta_{\nu \sigma}$
due to orthonormality of the eigenfunctions. Thus, 
$v_{\nu \sigma} = 0 \hspace{5pt}\forall \hspace{5pt} \nu\neq \sigma$.
Then Eq.~\ref{C_differential} admits the exact solution :
\begin{align}
\label{C_exact}
&C_{\alpha\nu}(t) = C_{\alpha\nu}(0)e^{-w_{\alpha \nu}t} + \varepsilon^2\frac{u_{\alpha\nu}^*+u_{\nu\alpha}}{w_{\alpha\nu}}(1-e^{-w_{\alpha \nu}t})
\end{align}
where $w_{\alpha\nu} = -i\omega_\alpha+\varepsilon^2 (-i f_{\alpha\alpha}^\Delta(\omega_\alpha)+f_{\alpha\alpha}(\omega_\alpha))+(\alpha \rightarrow \nu)^*$ and $u_{\nu\alpha}= F_{\nu\alpha}(\omega_\alpha)-i F_{\nu\alpha}^\Delta(\omega_\alpha)$. 
Note that the baths can still be at different temperatures and chemical potentials. So in this case, we have full time-dependent analytical closed form results for out of equilibrium correlation functions that hold for all values of $g$.

\subsubsection{Asymmetric couplings to baths : $\gamma_1 \neq \gamma_2$, $g\gg \frac{\varepsilon^2}{t_B}$ :}

Closed form analytical results are difficult to obtain for all $g$ when $\gamma_1 \neq \gamma_2$. But for $g\gg \frac{\varepsilon^2}{t_B}$, the analytical closed form for time dynamics can be found by solving Eq.~\ref{C_differential} perturbatively upto leading order in $\varepsilon$. The results are:
\begin{subequations}
\begin{align}
&N_{\alpha}(t) \simeq N_{\alpha}(0)e^{-2\varepsilon^2f_{\alpha \alpha}(\omega_{\alpha})t} + \frac{F_{\alpha \alpha}(\omega_{\alpha})}{f_{\alpha \alpha}(\omega_{\alpha})}(1-e^{-2\varepsilon^2f_{\alpha \alpha}(\omega_{\alpha})t}) \label{Nmt} \\
&C_{12}(t) \simeq C_{12}(0)e^{-w_{12}t} + \frac{i\varepsilon^2}{g}\Big [(F_{12}(\omega_2)\nonumber \\
&-i F_{12}^\Delta(\omega_2))(1-e^{-w_{1 2}t})  
+v_{21} (N_{1}(0)e^{-w_{1 2}t} -N_1(t)) \nonumber \\
&+ (1 \leftrightarrow 2)^* \Big] \hspace{5pt}   \label{Cnmt}
\end{align}
\end{subequations}
where $N_{\alpha}(t)=C_{\alpha\alpha}(t)$, and $w_{12} = -i\omega_1+\varepsilon^2 (-i f_{11}^\Delta(\omega_1)+f_{11}(\omega_1))+(1 \rightarrow 2)^*$. 

Note that, since we have already assumed $t \gg \tau_B$, the transient dynamics given by above equations can only be trusted if relaxation time of the system is much larger than $\tau_B$, viz., $\tau_B \ll {1}/{[\varepsilon^2 f_{\alpha \alpha}(\omega_{\alpha})]} \sim \frac{t_B}{\varepsilon^2}$, since ${f_{\alpha \alpha}(\omega_{\alpha})} \sim {t_B}$, from Eq.~\ref{Jform}, Eq.~\ref{qme} . Since $f_{\alpha \alpha}(\omega_{\alpha})>0$, it can be seen from  Eqs.~\ref{C_exact},\ref{Nmt},\ref{Cnmt} that the correlation functions approach  steady state values at time $t \gg {1}/{[\varepsilon^2 f_{\alpha \alpha}(\omega_{\alpha})]} \sim \frac{t_B}{\varepsilon^2}$, and the steady state values are independent of the initial state. So we can easily get steady state values of mode occupation $N_\alpha^{ss}$ and current between 1st and 2nd site $I_{1 \rightarrow 2}=-g\hspace{2pt} i\langle\hat{a}_1^\dagger \hat{a}_2-\hat{a}_2^\dagger \hat{a}_1\rangle = g\hspace{2pt} i\langle\hat{A}_1^\dagger \hat{A}_2-\hat{A}_2^\dagger \hat{A}_1\rangle = 2g$ $Im(C_{12})$ : 

\begin{subequations}
\begin{align}
&N_\alpha^{ss} \simeq  \frac{J_1(\omega_\alpha)n_1(\omega_\alpha)+J_2(\omega_\alpha)n_2(\omega_\alpha)}{J_1(\omega_\alpha)+J_2(\omega_\alpha)} \label{Nmss}\\
&I_{1 \rightarrow 2} \simeq \frac{\varepsilon^2}{2}\sum_{\alpha=1}^2 \frac{J_1(\omega_\alpha)J_2(\omega_\alpha)\big[n_1(\omega_\alpha) - n_2(\omega_\alpha)\big]}{ J_1(\omega_\alpha)+ J_2(\omega_\alpha)} \label{Cmnss}
\end{align}   
\end{subequations}
It is seen from the steady state equations that in equilibrium, i.e, when $n_1(\omega)=n_2(\omega)=n(\omega)$, $N_\alpha=n(\omega)$ and current is zero, which are the expected thermal values. Thus RQME shows proper thermalization and approach to steady state.

\subsection{Reduction to Lindblad form }
\label{lindblad}

The RQME is not in the Lindblad form. But it can be reduced to the Lindblad form by making certain further approximations. There are two popular forms of the
Lindblad equations using either the local operators $a_\ell$ or the eigenbasis 
operators $A_\nu$ [Eq.~\ref{Aop}]. We briefly discuss how they are obtained and their expected regimes of validity.

\subsubsection{Local Lindblad QME (LLQME) ( $g<\varepsilon$)  }
The local Lindblad equation for this system has the form 
${\partial \rho}/{\partial t} = i[\rho,\hat{\mathcal{H}}_S]+\varepsilon^2 \Big( \mathcal{L}_1^{LL}\rho+\mathcal{L}_2^{LL}\rho \Big) \label{LQME}$
where 
\begin{align}
\mathcal{L}_\ell^{LL}\rho =& J(\omega_0)e^{\beta_\ell(\omega_0-\mu_\ell)}n(\omega_0)(\hat{a}_\ell \rho \hat{a}_\ell^{\dagger} - \frac{1}{2}\{\hat{a}_\ell^{\dagger}\hat{a}_\ell,\rho\}) \nonumber \\
&+J(\omega_0)n(\omega_0)(\hat{a}_\ell^{\dagger} \rho \hat{a}_\ell - \frac{1}{2}\{\hat{a}_\ell \hat{a}_\ell^{\dagger},\rho\}) 
\end{align}
For $g<\varepsilon$, RQME Eq.~\ref{qme} can be reduced to LLQME  by expanding the non-unitary dissipative part about $g=0$ and keeping the first term. This is because, the dissipative part is already $O(\varepsilon^2)$, and since $g<\varepsilon$, higher order terms in $g$ will give higher order terms in $\varepsilon$, which we neglect in RQME treatment.  This amounts to putting $\omega_\nu = \omega_0$ in the dissipative part of Eq.~\ref{qme}. The same result is more conventionally obtained by considering the inter-site hopping term in the system Hamiltonian to be small and treating it as a part of the system-bath Hamiltonian while deriving the QME. This directly leads to the LLQME.  Thus LLQME is valid when $g<\varepsilon$. LLQME results for current and occupation for this problem have been derived in various papers \cite{GSA2013,Asadian2013}. The equilibrium condition $N_\alpha=n(\omega)$  is not obtained from LLQME, since for $g<\varepsilon$, each site interacts with its bath more strongly than with the other site, thereby thermalizing with its own bath. It clearly follows that if there was only one site, it would show thermalization. Thus the regime of validity of LLQME is too restrictive to show thermalization for a system with more than one non-interacting degrees of freedom. 

However, in interacting systems, it has been recently shown that even LLQME is capable of showing themalization \cite{Znidaric2010}.

\subsubsection{Eigenbasis Lindblad QME (ELQME) (~$g \gg \frac{\varepsilon^2}{t_B}, C_{12}^{ss}=0$~)}

The eigenbasis Lindblad equation for this system has the form 
$\frac{\partial \rho}{\partial t} = i[\rho,\hat{\mathcal{H}}_S]+\varepsilon^2 \Big(\mathcal{L}_1^{EL}\rho+\mathcal{L}_2^{EL}\rho \Big) \label{LQME}$
where 
\begin{align}
\mathcal{L}_\alpha^{EL}\rho =& (f_{\alpha\alpha}(\omega_\alpha) \mp F_{\alpha\alpha}(\omega_\alpha))(2\hat{A}_\alpha \rho \hat{A}_\alpha^{\dagger} - \{\hat{A}_\alpha^{\dagger}\hat{A}_\alpha,\rho \}) \nonumber \\
&+F_{\alpha\alpha}(\omega_\alpha)(2\hat{A}_\alpha^{\dagger} \rho \hat{A}_\alpha - \{\hat{A}_\alpha \hat{A}_\alpha ^{\dagger},\rho\})
\end{align}
Eq.~\ref{qme} is reduced to ELQME under rotating wave / secular approximation \cite{Spohn_Lebowitz1978,Dumcke_Spohn1979}, which amounts to neglecting $\alpha \neq \nu$ terms in the sum in Eq.~\ref{qme}. The rotating wave approximation assumes that the observation time $t \gg \frac{1}{g}$. On the other hand, to give the correct steady state, the QME must be valid for times shorter that time taken to reach steady state. This means that we need  the QME to be valid at times $t \lesssim \frac{t_B}{\varepsilon^2}$, since we have already seen that the time taken to reach steady state $\sim \frac{t_B}{\varepsilon^2}$. The two conditions are valid together if $g \gg \frac{\varepsilon^2}{t_B}$. This seems like a rather weak condition. 

However we note that, while these are the necessary conditions for obtaining ELQME, there is no guarantee that the resulting ELQME reproduces all physical observables accurately. 
In particular, neglecting $\alpha \neq \nu$ terms in the sum in Eq.~7 means ELQME has no terms connecting $\hat{A}_1$, $\hat{A}_2$. Therefore it gives $C_{12}^{ss}=0$, where $C_{12}^{ss}$ is the steady state value of $C_{12}$. This condition is of course valid only in equilibrium, where there is no current. Thus though rotating wave / secular approximation is a good approximation for equilibrium properties, it is a bad approximation in non-equilibrium. This point has also been succinctly discussed in a recent work \cite{Wichterich2007}.

However, ELQME still can be used to correctly obtain some non-eqilibrium results. For example, for $g\gg \frac{\varepsilon^2}{t_B}$ ELQME gives the same equation for $N_\alpha$ as Eq.~\ref{Nmt}, \ref{Nmss}. This result then can be used to obtain the correct current between the left bath and the system \cite{Esposito2007,Levy2014}. Thus ELQME suffers from a drawback that one is only able to compute the net current flowing between the two reservoirs in the NESS but not the current distributions in the system (e.g current flowing along two arms in a ring geometry). This also indicates a physical inconsistency of the ELQME formalism in non-equilibrium (see Appendix).

\begin{figure}[t]
\includegraphics[scale=.45]{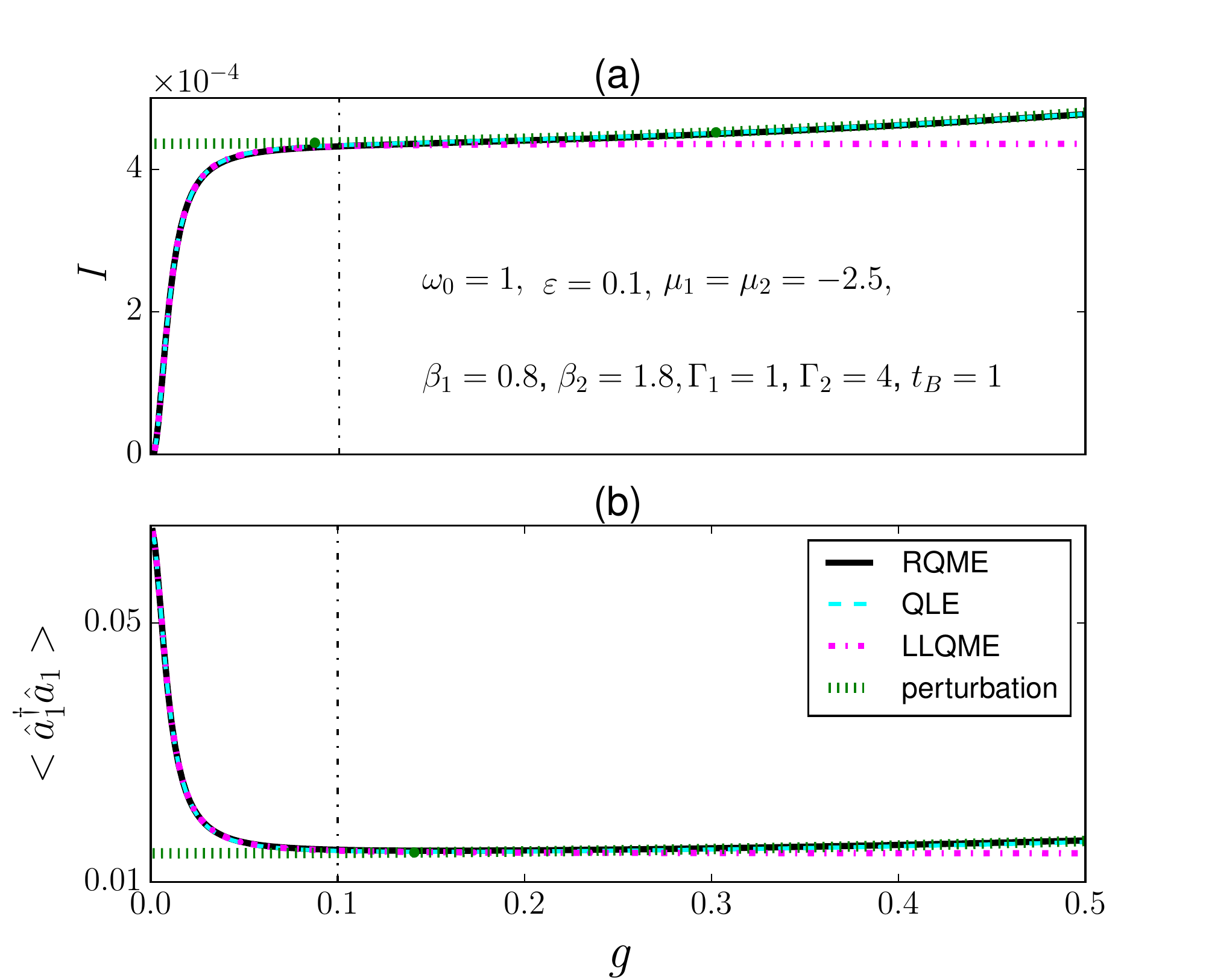} 
\caption{ (color online)  Bosonic model-steady state properties: the figure shows (a) particle current, (b) the occupation number in the left site, as a function of inter-site hopping $g$ for the $2$ site boson problem.  RQME shows near perfect agreement with exact results from QLE for all values of $g$, while LLQME and perturbation results [Eqs.~\ref{Nmss},\ref{Cmnss}] are valid in their respective limits. The vertical line marks the position of $g=\varepsilon$, below which LLQME is valid.  The parameter $\Gamma_{1,2}={2\gamma_{1,2}^2}/{t_B}$ is related to the system-bath couping [see Eq.~\ref{Jform}]. Current is measured in units of $\omega_0$ and all energy variables are measured in units of $\hbar\omega_0$.} \label{fig:one}
\end{figure}

\begin{figure}[t]
\includegraphics[scale=.45]{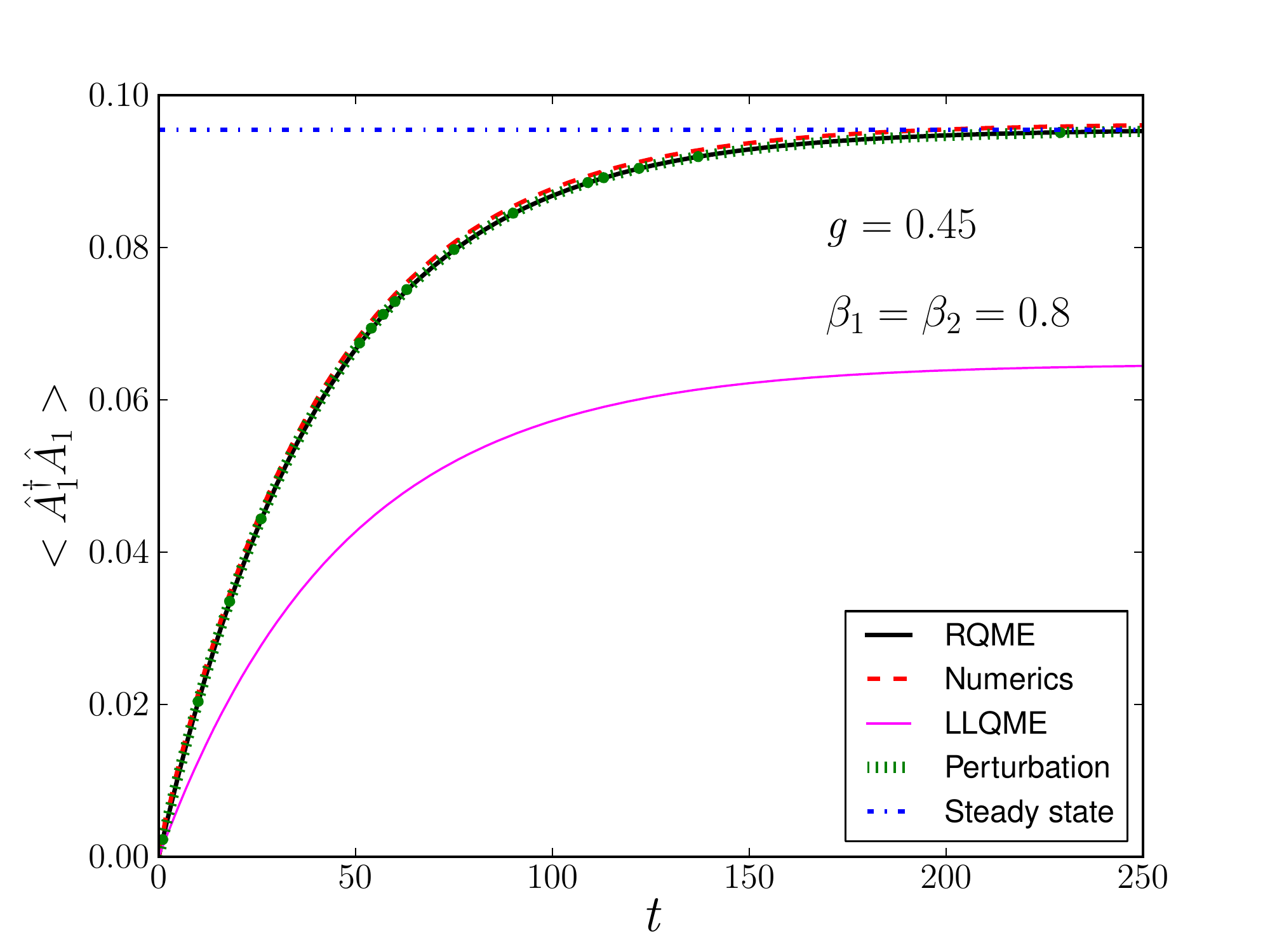} 
\caption{(color online) Bosonic model - thermalization: we show the time evolution of occupation of the lower energy mode $\hat{A}_1$, corresponding to energy $\omega_1=\omega_0-g$, in equilibrium for the $2$ site boson problem, starting from an empty system,  for $g=0.45$. The steady state (horizontal line) corresponds to the value of bose distribution $n(\omega_1) =[{e^{\beta( \omega_1-\mu)}-1}]^{-1}$ with the equilibrium bath temperatures and chemical potentials. LLQME does not show thermalization because it is not valid for $g>\varepsilon$, while  exact numerical results, RQME results and the perturbation result [Eqs.~\ref{Nmt}] match and show thermalization.    All parameters not explicitly specified are same as Fig.~(\ref{fig:one}). Current is measured in units of $\omega_0$ and all energy variables are measured in units of $\hbar\omega_0$. Time is measured in units of $\omega_0^{-1}$.} \label{fig:two}
\end{figure}  

\begin{figure}[t]
\includegraphics[scale=.45]{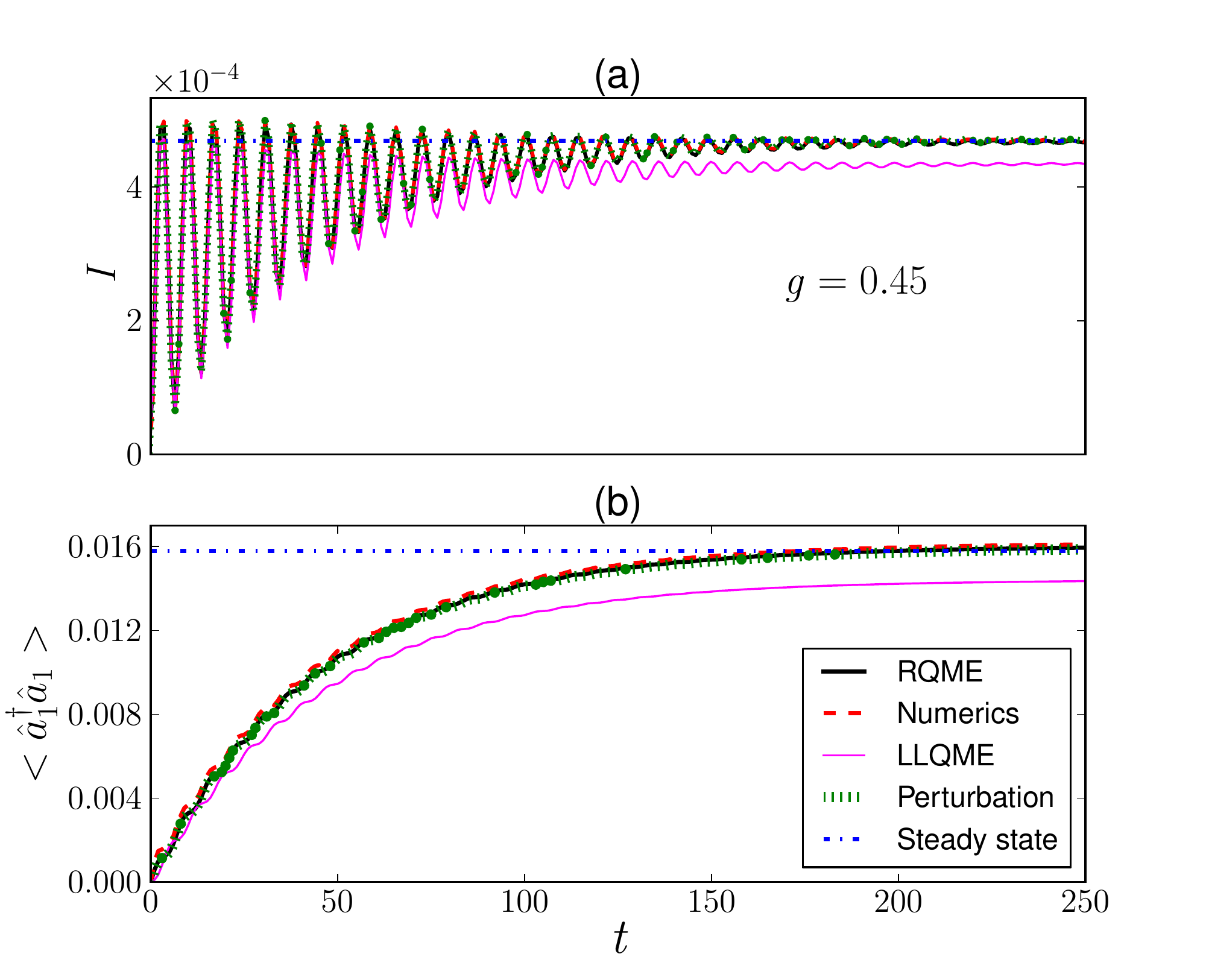} 
\caption{(color online) Bosonic model - non-equilibrium time dynamics: we show the time evolution of (a) particle current (b) occupation number of left site for the $2$ site boson problem, starting from an empty system,  for $g=0.45$. RQME results show good agreement with exact numerics. Since $g>\varepsilon$, LLQME does not match, while the perturbation result [Eqs.~(\ref{Nmt},\ref{Cnmt})] matches with the RQME. The horizontal line shows the exact steady state result obtained from QLE.  All parameters not explicitly specified are same as Fig.~(\ref{fig:one}). Current is measured in units of $\omega_0$ and all energy variables are measured in units of $\hbar\omega_0$. Time is measured in units of $\omega_0^{-1}$.} \label{fig:three}
\end{figure}

\subsection{Comparison of results from various methods and discussions}
\label{comp}

Finally we now present a detailed comparision of  results obtained using the 
various approaches, for both steady state and time-dependent properties. 
For the two site problem we consider the bosonic and fermionic versions and compute quantities such as the occupation number and particle current from site $1$ to site $2$. 
We again summarize the various approaches that we use:

(i) For steady  state properties, these are exactly computed using Eqs.~\ref{Iqle},\ref{occqle} following from the QLE approach.

(ii) Exact time dependent properties are  obtained  from the  numerical 
approach discussed in Sec.~(\ref{dynamics}).

(iii) The equations for correlation functions, Eqs.~\ref{C_differential}, are solved to obtain the predictions of RQME.  We also evaluate the perturbative solution of these equations given in Eqs.~\ref{Nmt},\ref{Cnmt}.

(iv) One can also write the equations for two-point correlations obtained from the Lindblad approach, and these are solved to obtain the predictions from LLQME.

(v) The ELQME approach cannot directly give the current inside the system. The predictions for the occupation number are the same as those from the perturbative solution of Eq.~\ref{Nmt}.   

We emphasize that all the approaches that we discuss are based on the same starting microscopic model of system and baths,  given by Eqs.~\ref{ham2S}, 
which lead to the bath spectral function Eq.~\ref{Jform}.

\begin{figure}[t]
\includegraphics[scale=.45]{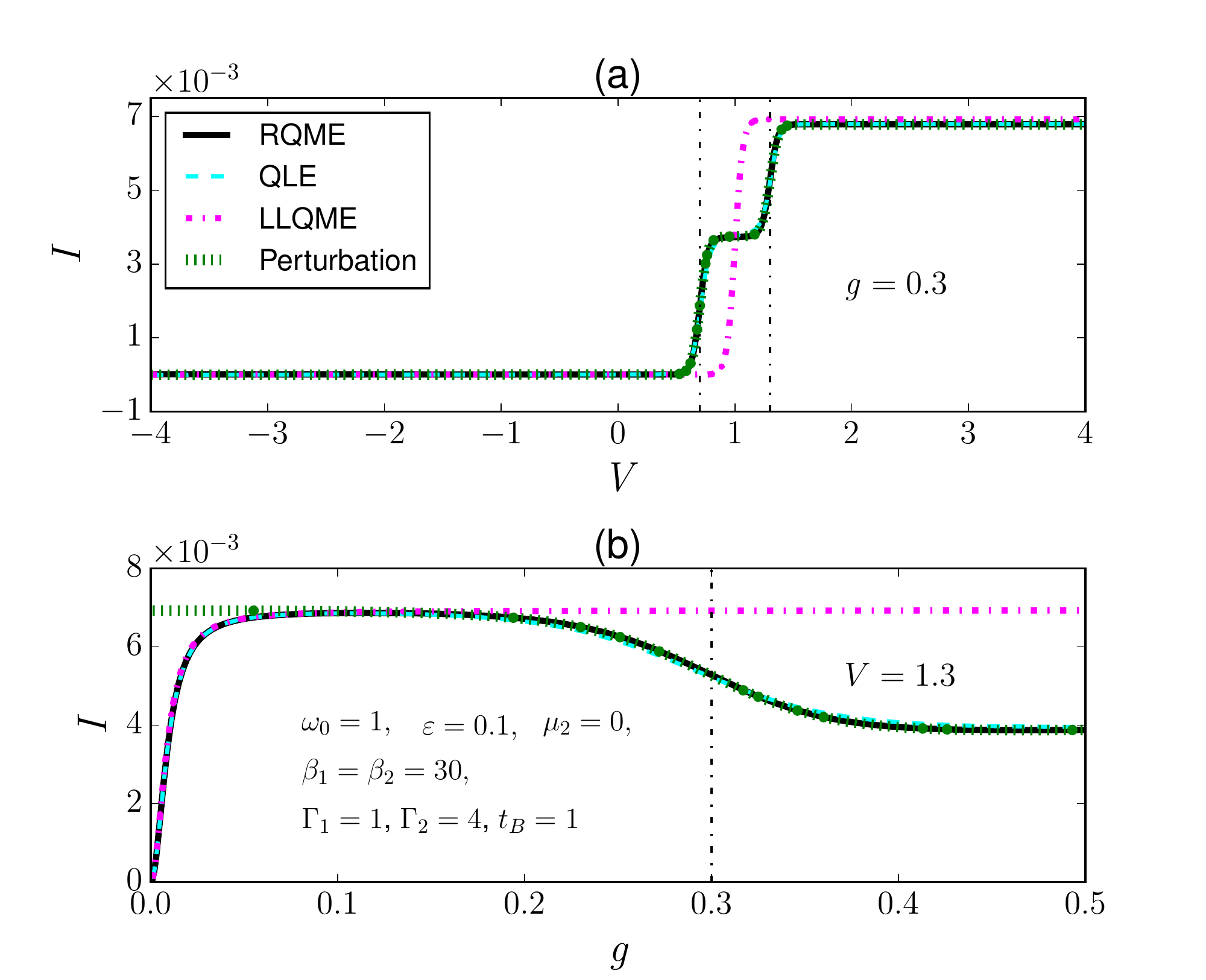}  
\caption{\footnotesize (color online) Fermionic model - steady state properties: the plot shows  (a) particle current vs voltage (b) particle current vs inter-site hopping in the $2$ site fermion problem. The vertical lines correspond to positions where potential difference $V=\mu_1-\mu_2=\omega_\alpha$, where $\omega_\alpha$ are the system energy levels. RQME shows near perfect agreement with exact results from QLE for all values of $g$, while LLQME and perturbation results [Eqs.~\ref{Nmss},\ref{Cmnss}] are valid in their respective limits. The graphs demonstrate the effect of conductance quantisation at low temperatures. The parameter $\Gamma_{1,2}={2\gamma_{1,2}^2}/{t_B}$. Current is measured in units of $\omega_0$ and all energy variables are measured in units of $\hbar\omega_0$.} \label{fig:five}
\end{figure}

For the bosonic case, the steady state results for current and occupation number 
are shown in Fig.~(\ref{fig:one}),  and results from time-dynamics  in Fig.~(\ref{fig:two},\ref{fig:three}).  For the time dynamics, for the results presented here, the initial condition corresponds to no particles inside the system and baths in equilibrium at different $\mu$ and $T$. But we have tested with other initial conditions like a finite number of particles in the system and random initial values of the correlation functions.  Our following observations are true for such generic initial conditions for the system. 
For the fermionic case, the results for the steady state current as a function of voltage difference and of the intra-system coupling $g$, are shown in Fig.~(\ref{fig:five}).  In all cases, the system-bath coupling is chosen to be asymmetric, i.e, $\gamma_1 \neq \gamma_2$. 

Our most important observation is that RQME results obtained from the solution of Eq.~\ref{C_differential} agree very well with exact results from QLE and numerics, for all values of $g$ for steady state, as well as for long time dynamics. The LLQME agrees well for $g<\varepsilon$ as expected. In Fig.~(\ref{fig:two}), we show that the system indeed thermalizes in equilibrium, and this is perfectly captured by RQME, and not by LLQME since it is invalid for $g>\varepsilon$.  For $g \gg \frac{\varepsilon^2}{t_B}$, 
our analytical closed form perturbation results match quite well with the exact results, showing correct approach to steady state and thermalization. Therefore, we conclude that RQME gives the correct physics under Born-Markov approximation.

Apart from validating the RQME description, we also observe interesting physical trends.  The boson problem may be realised in bosonic cold atom experiments or in optical cavity experiments with suitable choice of parameters and spectral functions. We see in Fig.~\ref{fig:one} that steady state  properties have markedly different behaviour depending on whether $g<\varepsilon$ or $g>\varepsilon$. For $g<\varepsilon$, the current increases rapidly, but after that there is a slow increase in current. Also, for $g<\varepsilon$ the occupation of the left cavity becomes minimum when $g \simeq \varepsilon$. However, beyond this point, the occupation of the left cavity increases slowly with increase in tunnelling probability $g$. These trends may be experimentally observed. However, these trends depend on the choice of the bath spectral function $J_\ell(\omega)$. For example, for optical cavity experiments, the commonly used Ohmic dissipation $J_\ell(\omega) \propto \omega$ will give  slow decrease of current with $g$ for $g>\varepsilon$, still showing a markedly different behaviour from $g<\varepsilon$ case.  Here we microscopically derived $J_\ell(\omega)$ assuming a microscopic model of the bath.

The fermionic system of two-sites may be experimentally realized in non-interacting quantum dots or in fermionic cold atom experiments. The current versus voltage plot of the fermionic system shows effect of conductance quantization, which is observed experimentally \cite{Krinner2015}. The current versus hopping $g$ plot shows suppression of current after a value of $g$. These observations can be explained as follows. The two site system has two eigen-energy levels of energy $\omega_0-g$ and $\omega_0 + g$ respectively. In Fig.~\ref{fig:five}, the right bath is held at zero chemical potential while the chemical potential of the left bath, $\mu_1$, is varied. When $\mu_1 \ll \omega_0-g$, no fermion from the left bath has the energy to enter the system. So there is no flow of current. When $\omega_0-g \ll \mu_1 \ll \omega_0+g$, the fermions can hop through the system via the lower energy level. So a finite current flows through the system. When $\mu_1 \gg \omega_0+g$, fermion transport through system can occur through both energy levels. Since all system levels are now participating in transport, increasing $\mu_1$ beyond this point does not affect current any more. For similar reason, suppression of current occurs when $g \gg \mu_1-\omega_0$ in current versus $g$ plot. These observations can be easily obtained for larger systems also.       

The NESS for the fermionic problem has  been solved earlier by a RQME method, but in the limit of negligible Fock space coherences \cite{Harbola2006}, which give the results identical to $g \gg \frac{\varepsilon^2}{t_B}$ results Eqs.~\ref{Nmss},\ref{Cmnss}.  However, their  \cite{Harbola2006}  method of solution cannot easily be generalized to treat the bosonic version of the problem. 

One important issue with the Redfield operator is that it is not completely positive.  In a recent work \cite{Argentieri2014}, it has been shown to lead to negative entropy production in a specific system. 
In this manuscript we find RQME to be giving the correct Born-Markov approximated results for correlation functions (closely related to physical observables), both equlibrium and time dynamics. But the situation regarding non-positivity of Redfield operator is not completely clear and requires further investigation.

\section{Conclusions}
\label{conclusions}
Quantum master equations derived under Born-Markov approximation are widely used to describe open quantum systems. But the validity of such approximate QMEs have not been, to our knowledge, rigoursously checked before. In this work, by studying a simple system of non-interacting bosons/fermions and comparing results from microscopically derived Redfield QME under Born-Markov approximation with exact results, we show that the Redfield QME indeed gives very accurate results in a wide parameter range. Moreover, we observe that, rotating wave/secular approximation that are often done to reduce the Redfield QME to the eigenbasis Lindblad form is inappropriate in a non-equilibrium setting. On the other hand, the small hopping approximation that is often employed to reduce Redfield QME to local Lindblad form is very restrictive for non-interacting particles. These observations are consistent with previous results \cite{BauerNotes, Levy2014, Wichterich2007}. So one is required to work with the full Redfield QME to obtain correct physics in a non-equilibrium system of non-interacting bosons/fermions.  

One of the reasons a Redfield equation is often reduced to the Lindblad form is easy numerical treatment. The Redfield equation is in general much more difficult to solve than the Lindblad equations. However, for the case of non-interacting bosons/fermions in an arbitrary lattice in any dimension, here, we presented a complete set of linear differential equations (Eq.~ \ref{C_differential}) for two-point correlation functions which can be easily solved numerically. Eq.~\ref{C_differential}, therefore, gives a numerical way to directly calculate time-dynamics of out-of-equilibrium two-point correlation functions in an arbitrary system of non-interacting particles in a lattice. Various physical observables can be directly calculated from such correlation functions. 

For a two site system we derived closed form analytical expressions for out-of-equilibrium time dynamics of two-point correlation functions (Eqs.~\ref{C_exact},\ref{Nmt}, \ref{Cnmt}). These analytical results explicitly show that the correlation functions approach a steady state value independent of initial conditions, and these steady state values give expected thermalization in equilibrium. They also match with numerically obtained exact results from other methods independent of any approximate master equation, thereby validating the Redfield approach.  Various physical quatities like site-occupation and current calculated from the correlation functions show interesting experimentally observable trends which are seen in parameter regimes beyond the regime of validity of Lindblad methods. These trends may be seen in experiments with cavity QED, cold atom, quantum dots etc. 
Needless to mention, recent cutting-edge technologies available can be exploited to engineer Hamiltonian and reservoirs such that the discrepancy between Lindblad methods and exact results can be made visible.  

The non-positivity of the Redfield operator remains an important issue and needs further work. However, this does not seem to be a problem in reproducing correct physical steady-state properties as well as correct long time dynamics in our system.

We re-emphasize that the main conclusion from our present work is that it is indeed justified to adopt the RQME in order to go beyond the limitations of the Lindblad methods.
Future work involves RQME with similar rigorous treatment of reser-
voirs for open quantum non-linear (interacting) systems where exact
methods are not available, for e.g., Jaynes−Cummings, spin-boson,
open XY spin chains, Dicke type models  \cite{Rochetal2014,Aronetal2014,ucb2015,kulkarniprl,
esslinger_science,tilman_2013,porras,rhqd,rhqd1,rep2,Aron2016}). Such a RQME approach could unravel interesting out-of-equilibrium many body phenomena that may be missed by conventional Lindblad-type approaches.

\textit{Acknowledgements:}
We acknowledge fruitful discussions with R. Vijay, Y. Dubi, K. Saito, M. Esposito, J. L. Lebowitz, D. Huse and H. Spohn. M.K gratefully acknowledges support from the Professional Staff Congress of the City University of New York award No. 68193-0046. A.D. acknowledges support from Indo-Israel joint research project F.No. 6-8/2014(IC).  We thank the hospitality of Raman Research Institute and International Centre for Theoretical Sciences of the Tata Institute of Fundamental Research, Bengaluru during the workshop on Statistical Physics where several interesting discussions took place. We would also like to thank the hospitality of the International Centre for Theoretical Sciences of the Tata Institute of Fundamental Research, Bengaluru during the programme ``Non-equilibrium statistical physics'' where this work culminated.

\section{\label{appendix}Appendix}

\textit{Derivation of Redfield QME:} Here, we briefly outline the steps involved in obtaining our QME Eq.~\ref{qme} from Eq.~\ref{gen_master}. We first go to the eigenbasis of the system through the transformation: $\hat{a}_{\ell} = \sum_{\alpha=1}^N c_{\ell \alpha} \hat{A}_\alpha$. Then to use Eq.~\ref{gen_master}, we need to go to the interaction picture representation,
$\hat{a}_{\ell}^I = \sum_{p=1}^N c_{\ell \alpha} \hat{A}_p e^{-i \omega_\alpha t}$,  $\hat{B}_r^{\ell I}(t) = \hat{B}_r^\ell e^{-i \Omega_r^{\ell} t}$. Next, substituting into Eq.~\ref{gen_master} and using the bath correlations Eq.~\ref{initial_bath_corr} and bath spectral functions Eq.~\ref{J}, and going back to Schroedinger picture we obtain :
\begin{equation}
\label{qme_deriv}
\begin{split}
&\dot{\rho}(t) = i[\rho, \hat{\mathcal{H}}_S] - \varepsilon^2\sum_{\alpha,\nu,\ell=1}^N c_{\ell \alpha}^* c_{\ell \nu}\int_{\Omega_{min}^{\ell}}^{\Omega_{max}^{\ell}} \frac{d\omega}{2\pi} \Big[ \Big\{[\rho(t) \hat{A}_\nu, \hat{A}_\alpha^{\dagger}]  \\
&+[\hat{A}_\alpha^{\dagger}, \hat{A}_\nu \rho(t)] e^{\beta_\ell(\omega-\mu_\ell)} \Big\} J_\ell(\omega) n_\ell(\omega)  \int_0^t d\tau e^{i(\omega - \omega_\nu)\tau} \Big] \\&+\hspace{2pt}  h.c.
\end{split}
\end{equation}
where $\Omega_{min}^{\ell},\Omega_{max}^{\ell}$ are the minimum and the maximum energy levels of the bath coupled to the $\ell$th site. Now, assuming $t \gg \tau_B$, where $\tau_B$ is the relaxation time scale of the bath, we replace the upper limit of the time integral in the above equation by $t \rightarrow \infty$ to obtain our QME Eq.~\ref{qme}. Even though this assumption is guaranteed to give the correct steady state, an estimate of $\tau_B$ is required to ensure validity of the transient dynamics from our QME. $\tau_B$, of course, depends on the model of bath. 

\textit{A bath model in wide-band limit:} The model of bath enters the calculation through the bath spectral function $J_{\ell}(\omega)=2\pi \sum_r  \mid \kappa_{\ell r} \mid^2 \delta(\omega - \Omega_r^\ell)$.  Note that since in our derivation of RQME, the system couples to the eigenmodes of the baths, $\kappa_{\ell r}$ are proportional to the eigenfunctions of the bath Hamiltonian. Because of infinite degrees of freedom, the energy spectrum of the bath can be considered continuous. For our case of the Hamiltonian in Eq.~\ref{ham2S}, bath eigen-energies  are $\Omega(q_\ell) = -2t_B \cos q_\ell$ and  $\kappa(q_\ell)=\gamma_\ell\sqrt{\frac{2}{\pi}}\sin q_\ell$, with $0 \leq q_\ell \leq \pi$. Thus, 
\begin{equation}
\begin{split}
J_{\ell}(\omega) =& 4\gamma_\ell\int_0^\pi dq_\ell \sin^2 q_\ell \hspace{3pt}\delta(\omega+2t_B \cos q_\ell) \\
& = \frac{2\gamma_\ell^2}{t_B} \sqrt{1-\frac{\omega^2}{4t_B^2}}
\end{split}
\end{equation}
We also need to the following result $ \mathcal{P} \int \frac{}{}\frac{d\omega^{\prime} J_{\ell}(\omega^{\prime})}{2\pi(\omega^{\prime}-\omega)}=-\frac{\gamma_\ell^2\omega}{2t_B^2}$ to calculate $f_{\alpha\nu}^{\Delta}(\omega)$ [in Eq.~\ref{C_differential}]. $F_{\alpha\nu}^{\Delta}(\omega)$ cannot be written in a simple closed form and is calculated numerically.   The functions $f_{\alpha\nu}(\omega),F_{\alpha\nu}(\omega)$ can be written down in matrix form as :

\begin{align}
&
f(\omega) = \frac{1}{4}
\begin{bmatrix}
J_1(\omega) + J_2(\omega) & J_1(\omega) - J_2(\omega) \\
J_1(\omega) - J_2(\omega) & J_1(\omega) + J_2(\omega)
\end{bmatrix}
\\
&
F(\omega) = \nonumber\\
&\frac{1}{4}
\begin{bmatrix}
J_1(\omega)n_1(\omega) + J_2(\omega)n_2(\omega) & J_1(\omega)n_1(\omega) - J_2(\omega)n_2(\omega) \\
J_1(\omega)n_1(\omega) - J_2(\omega)n_2(\omega) & J_1(\omega)n_1(\omega) + J_2(\omega)n_2(\omega)
\end{bmatrix}
\end{align} 

\textit{Estimate of $\tau_B$:} Now, for this bath model we estimate $\tau_B$. Looking at Eq.~(\ref{qme_deriv}), we find integrals of the type $\mathcal{I}_1(\omega_\nu,t)=\int_0^t d\tau e^{-i\omega_\nu\tau} \mathcal{I}_2(\tau)$ with $\mathcal{I}_2(\tau)$ given by :
\begin{equation}
\mathcal{I}_2(\tau) = \int_{\Omega_{min}^{\ell}}^{\Omega_{max}^{\ell}} \frac{d\omega}{2\pi}J_\ell(\omega) n_\ell(\omega)e^{i\omega\tau} 
\end{equation}
The time $\tau_B^{(\ell)}$ at which $\mathcal{I}_2(\tau)$ decays is the relaxation time of the $\ell$th bath.  If the bandwidth of the bath is large enough, $\mathcal{I}_2(\tau)$ is like a Fourier transform of $J_\ell(\omega) n_\ell(\omega)$, and hence the $\tau_B^{(\ell)}$ depends on the spread of $J_\ell(\omega) n_\ell(\omega)$. At low temperatures the spread will be $\sim 1/\beta_\ell$. Hence $\tau_B^{(\ell)} \sim \beta_\ell$.   

As mentioned in the manuscript, the transient dynamics from our QME is valid only if the system relaxation time is much greater than $\tau_B$. From Eq.~\ref{Nmt},\ref{Cnmt}, we see that system relaxation time is $\sim \frac{1}{\varepsilon^2 f_{\alpha \alpha}(\omega_{\alpha})}$. Thus, for validity of our transient dynamics,
\begin{equation}
\tau_B \ll \frac{1}{\varepsilon^2 f_{\alpha \alpha}(\omega_{\alpha})} \hspace{5pt} \forall \hspace{5pt} \alpha \in {1,2,..,N}
\end{equation}
For our $2$-site case, $f_{\alpha \alpha}(\omega_{\alpha}) \sim t_B^{-1}$. Thus, the condition for validity of transient dynamics for the $2$-site case becomes,
\begin{equation}
\beta_\ell \ll \frac{t_B}{\varepsilon^2} \hspace{5pt} \forall \hspace{5pt} \ell=1,2
\end{equation}
This condition is clearly satisfied by our choice of parameters.

\textit{Problem with ELQME:}
In the ELQME, problems arise in the definitions of NESS current. For our Hamiltonian Eq.~ \ref{ham2S}, current can be derived from the following equations :
\begin{align}
&\frac{d\langle\hat{a}_1^\dagger \hat{a}_1\rangle}{dt} = I_{B^{(1)} \rightarrow 1} - I_{1 \rightarrow 2} \label{curr_def1}\\
&\frac{d\langle\hat{a}_1^\dagger \hat{a}_1+\hat{a}_2^\dagger \hat{a}_2\rangle}{dt} = I_{B^{(1)} \rightarrow 1} - I_{2 \rightarrow B^{(2)}} \label{curr_def2}
\end{align}   
 
where $I_{B^{(1)} \rightarrow 1}$ is the current between left bath and left system site, $I_{1 \rightarrow 2}$ is the current between left and right system sites and $I_{2 \rightarrow B^{(2)}}$ is the current between right site and right bath.
Note that the expression for $I_{1 \rightarrow 2}$ is same from all three approaches RQME, LLQME, ELQME  because it comes from the non-dissipative part of the QME. On the other hand, depending on whether the approach is RQME or LLQME or ELQME, the expressions for $I_{B^{(1)} \rightarrow 1}$ and $I_{2 \rightarrow B^{(2)}}$ are different as they come from the dissipative part. In NESS all three currents defined above are equal. This is true for RQME and LLQME. But ELQME gives $I_{1 \rightarrow 2}=0$ even in NESS  while giving an non-zero current for $I_{B^{(1)} \rightarrow 1}$. In fact, Eq.~\ref{curr_def1}, from ELQME, becomes of the form $\frac{d<\hat{a}_1^\dagger \hat{a}_1>}{dt} = I_{B^{(1)} \rightarrow 1} - I_{1 \rightarrow B^{(2)}}$, where a fictitious $I_{1 \rightarrow B^{(2)}}$ current from left site to right bath appears which is completely unphysical as there is no direct connection between left site and right bath. However, if Eq.~\ref{curr_def2} is used, then ELQME gives the same result as obtained from the RQME in the limit $g\gg (\varepsilon^2 /t_B)$. This is because $I_{B^{(1)} \rightarrow 1}$ obtained from Eq.~\ref{curr_def2} depends only on $N_1,N_2$ which are correctly given by ELQME when $g \gg (\varepsilon^2 /t_B)$.  Thus, though ELQME is not physically self consistent, this trick can be used to obtain the correct current in our setup, as done in various places \cite{Esposito2007,Levy2014}. However, this trick will not work in cases with different geometries. For example, if two sites of a ring are connected to two different baths, ELQME will not be able to give current flowing in the two arms.

\textit{The Gibbs state:}
In equlibrium, the Gibbs state is defined as $\rho_{Gibbs} = \frac{e^{-\beta(\hat{\mathcal{H}}_S-\mu \hat{\mathcal{N}}_S)}}{Tr(e^{-\beta(\hat{\mathcal{H}}_S-\mu \hat{\mathcal{N}}_S)})}$ where $\hat{\mathcal{N}}_S = \sum_{\alpha=1}^N \hat{A}^{\dagger}_\alpha \hat{A}_\alpha$. Note that, contrary to as often reported in literature, the Gibbs state is not a stationary state of the our RQME (Eq.~\ref{qme}) in equilibrium. This is because we do not neglect the principal value terms $f_{\alpha\nu}^{\Delta}(\omega)$ and $F_{\alpha\nu}^{\Delta}(\omega)$ as is often done (For example, \cite{Wu2011,Harbola2006}).  We find no reason to neglect them as the terms are neither small, nor do they contribute to only shifting the system energy levels by a constant amount. For $g \gg (\varepsilon^2/t_B)$, the equilibrium steady state, which can be computed from the two point correlation functions, has $O(\varepsilon^2)$ corrections over the Gibbs state. So, in this case, RQME gives the correct thermalization in the following sense :
\begin{equation}
\lim_{\varepsilon\rightarrow 0} \lim_{t\rightarrow \infty} \rho(t) = \rho_{Gibbs}, \hspace{3pt} \forall \hspace{3pt} \beta_1=\beta_2=\beta, \hspace{3pt} \mu_1=\mu_2=\mu
\end{equation}      
where the order of limits in important. ELQME, on the other hand, has the Gibbs state as its stationary state in equilibrium.  

\bibliographystyle{unsrt}


\end{document}